%% file: Gebran.tex
\documentclass[twoside]{sfm}

\input{sf.def}

\begin{document}

\def\llm{{\sc LLmodels}}
\def\atl{{\sc ATLAS9}}
\def\aatl{{\sc ATLAS12}}
\def\starsp{{\sc STARSP}}
\def\aur{$\Theta$~Aur}
\def\logg{\log g}
\def\tauros{\tau_{\rm Ross}}
\def\kms{km\,s$^{-1}$}
\def\bz{$\langle B_{\rm z} \rangle$}
\def\degr{^\circ}
% journals
\def\aaps{A\&AS}
\def\aap{A\&A}
\def\apjs{ApJS}
\def\apj{ApJ}
\def\rmxaa{Rev. Mexicana Astron. Astrofis.}
\def\mnras{MNRAS}
\def\actaa{Acta Astron.}
\newcommand{\Tef}{T$_{\rm eff}$~}
\newcommand{\Vt}{$V_t$}
\newcommand{\CC}{$^{12}$C/$^{13}$C~}
\newcommand{\CDC}{$^{12}$C/$^{13}$C~}

\input{Geb.tex}

\end{document}

%% file: Geb.tex
\pagebreak

\thispagestyle{titlehead}

\setcounter{section}{0}
\setcounter{figure}{0}
\setcounter{table}{0}

%%%%%%%%%%%%%%%%%%%%%%%%%%%
% !!!
\markboth{Gebran et al.}{Microturbulence in A/F Am/Fm stars}

\titl{Microturbulence in A/F Am/Fm stars}{Gebran M.$^1$, Monier R.$^{2,3}$, Royer F.$^4$, Lobel A.$^5$, and Blomme R.$^5$}
{$^1$Department of Physics \& Astronomy, Notre Dame University -- Louaize, Lebanon, email: {\tt mgebran@ndu.edu.lb} \\
 $^2$LESIA/CNRS UMR 8109, Observatoire de Paris -- Universit\'e Paris Denis Diderot, 5 place Jules Janssen, 92190 Meudon, France\\
 $^3$Laboratoire Lagrange, Universit\'e de Nice Sophia Antipolis, Parc Valrose, 06100 Nice, France\\
 $^4$GEPI/CNRS UMR 8111, Observatoire de Paris -- Universit\'e Paris Denis Diderot, 5 place Jules Janssen, 92190 Meudon, France\\
$^5$ Royal Observatory of Belgium, Ringlaan 3, B--1180 Brussels, Belgium
}

\abstre{
A programme to observe all A dwarfs in open clusters brighter than V=6.5 mag of various ages and in the field was initiated several years ago. In this work we present the current status of microturbulent velocity for A and F dwarfs. We have performed high resolution high signal-to-noise spectroscopy of stars well distributed in mass along the Main Sequence.
Microturbulent velocities are derived iteratively by fitting grids of synthetic spectra  calculated
 in LTE to observed spectra of 61 A field stars, 55 A and 58 F in open clusters (Pleiades, Coma Berenices, Hyades and the Ursa Major moving group). We compared our results to recent works and found a good agreement with their analytical formulation for the standard microturbulence. Our results show a broad maximum for microturbulent velocities in the range A5V to about A9V and a decrease (to $\sim 1$ km/s) for cooler and hotter stars as indicated in Smalley (2004).We also present a comparison to preliminary science results of Lobel et al. (2013) for the Gaia-ESO Public Spectroscopic Survey.
}

\baselineskip 12pt

\section{Introduction and Program stars}

The microturbulent velocity $\xi_t$, usually interpreted  as a microscale non-thermal velocity, has a significant effect on the emergent flux of A and F stars. It is used as a free fitting parameter in order to obtain equal abundances from lines of different equivalent width.\\
The derived values of $\xi_t$ are used to establish a relation between $\xi_t$ and $T_{\rm{eff}}$ for A and F stars (Coupry \& Burkhart (1992), Edvardsson et al. (1993), Gray et al. (2001) , Smalley (2004), Gebran \& Monier (2007), Takeda et al. (2008), \ldots). In this work, we merge results from our previous own work on A and F stars in open clusters with a recent survey of slowly rotating field A stars and new microturbulent velocity determinations achieved in the frame of the Gaia ESO Public Spectroscopic Survey. (GES PIs: G. Gilmore \& S. Randich). Note that these GES results are not yet official. The stars analysed in the present work are listed in the following table\ref{tab1}. AURELIE, ELODIE, and SOPHIE spectrographs are mounted on the telescopes at the Observatoire Haute-Provence (OHP) while GIRAFFE/FlAMES are mounted on the Very Large Telescope (VLT) at the European Southern Observatory (ESO).
\begin{table}
\small
\begin{tabular}{|c|c|c|}
\hline
\hline
\# of stars  & Spectrograph & Comments \\
\hline
16 A, 5 F in Pleiades& ELODIE/SOPHIE & Gebran \& Monier 2008 \\
11 A, 15 F in Coma& AURELIE/ELODIE & Gebran et al. 2008  \\
16 A, 28 F in Hyades& AURELIE/SOPHIE & Gebran et al. 2010 \\
12 A, 10 F in Ursa& AURELIE & Monier 2005 \\
61 A Field& ELODIE/SOPHIE, & Royer et al. in prep.\\
	&ECHELEC		&	Royer et al. (2002)\\
133 A \& early-B  in NGC 3293& Giraffe, & Gilmore, Randich, et al. (2012)\\
			& FLAMES	&		\\
\hline
\hline

\end{tabular}
\caption{Data of the observations.}
\label{tab1}
\end{table}

\section{Method}\label{method}
\textbf{For Pleiades, Coma, Hyades, Ursa Major, and field stars:}\\
Effective temperatures $T_{\rm{eff}}$ and surface gravities $\log g$ were determined using the new version of Napiwotzki et al.'s (1993) UVBYBETA code using Str\"omgren  uvby$\beta$ photometry as input. The errors on $T_{\rm{eff}}$ and $\log g$ are estimated to be $\pm125$ K and $\pm$0.2 dex respectively. LTE model atmospheres were computed using Kurucz's ATLAS9 (Kurucz, 1992) and ATLAS12 (Kurucz 2005) code, assuming a plane parallel geometry, hydrostatic equilibrium, radiative equilibrium and depth independent microturbulence (in ATLAS, $v_{turb}$, is connected to $P_{turb}$, the turbulent pressure and is equal to $1/2 \rho v_{turb}^2$ and has the same order of magnitude as the $\xi_t$ we derive by fitting but it is not a priori the same thing. There is no physical definition of $\xi_t$). The line list was built from Kurucz's gfall.dat (http://kurucz.harvard.edu), the oscillator strengths were changed using more accurate values from literature, from VALD (http://ams.astro.univie.ac.at/vald/) and from the NIST database (http://www.nist.gov).  The synthetic spectra were computed using Takeda's (1995) automated iterative procedure. 
The microturbulent velocity $\xi_t$, rotation velocity $v_e\sin i$ and the abundances are simultaneously determined by fitting line profiles. The lines used for fitting are the MgII triplet at 4481 \AA\ and the unblended FeII lines between 4491.405 \AA\ and 4522.634 \AA. The iron lines are more sensitive to rotational velocity while the MgII triplet is sensitive to both parameters ($\xi_t$, $v_e \sin i$). Combining the results of these lines, we deduce $v_e \sin i$ and $\xi_t$ of the star. These values are fixed later and used for abundance determinations.\\

\textbf{For NGC3293:}\\
The Astrophysical Parameters (APs) have been derived by Alex Lobel for 133 A- and early B-type stars observed for the Gaia ESO Survey in the galactic young open cluster NGC 3293.
% Multi-fibre VLT-Giraffe spectra were observed in 2012 with R$\simeq$20\,000 in various HR gratings.
Synthetic spectra have been computed under the LTE assumption using the code {\sc Scanspec} (see {\tt alobel.freeshell.org/scan.html}). The line $\log gf$-values are from {\sc SpectroWeb} (Lobel 2011). The Mg~{\sc ii} $\lambda$4481 triplet lines are very temperature sensitive in A-type stars and we use the equivalent width (EW) to obtain a first estimate of $T_{\rm{eff}}$. Synthetic spectra were computed using a grid of Kurucz's ATLAS9 model atmospheres using the opacity distribution functions of Castelli \& Kurucz (2003). The values of $T_{\rm{eff}}$ and $\log g$ are varied by 250 K and 0.5 dex. The EW of the Mg~{\sc ii} line at 4481.24 \AA have been measured after rectification of the observed spectrum to a local continuum. The model $T_{\rm eff}$ and $\log g$-values are varied in combination with $\xi_t$ (
in steps of 0.5~$\rm km\,s^{-1}$) until the measured EW is reproduced. It yields a series of initial models we utilize for calculating the theoretical spectrum around the H$\delta$ and/or H$\alpha$ line. The extended damping wings of the H Balmer lines are very sensitive to $\log g$, and we perform detailed fits to the line shape and depth for selecting the best matching model from the initial model series.

The second step of the APs determination procedure performs a detailed synthesis calculation between 4503 \AA\,- 4580 \AA\, and 4030 \AA\,- 4070~\AA. The automated abundance iterations are stopped for an iron abundance value that correctly fits the central core depths of the Fe~{\sc i} and Fe~{\sc ii} lines after rotational broadening. The iterations vary the $\xi_t$-value (typically by less than $\pm$0.5 to $\pm$1 $\rm km\,s^{-1}$) in order to minimize the difference between the abundance values obtained from the Fe~{\sc ii} line and the average of both Fe~{\sc i} lines. Hence, the $\xi_t$-value is consistent with the Fe~{\sc i}-Fe~{\sc ii} iron-ionization balance.

\section{Results}\label{res}
%\begin{center}
%\vskip 0.3cm
%\begin{figure}[b!]
%\includegraphics[scale=0.35]{Gebran/fig1.eps}
%\caption{Example of a fit of an observed spectrum of the A1m star HD72660 with a synthetic one calculated using Takeda's Procedure.}
%\label{hd72660}
%\end{figure}
%\end{center}
%As an example of the adjustment of an observed spectrum of an A star, we show in Fig.~1 an overplot of a synthetic spectrum of the A1m star HD72660 over the observed one. The plot corresponds to the calculation that minimises the chi-square between the synthetic and the observed spectra. The displayed region between 4475 and 4525 \AA\ is the one used to derive $v_e\sin i$ and $\xi_t$.

%The derived microturbulent velocities ($\xi_t$) are plotted as a function of the effective temperature ($T_{\rm{eff}}$) and surface gravity ($\log g$) in Fig.~1. 
In the left panel of Fig.~1 we display the values of $\xi_t$ for A and F stars (filled circles) member of the Pleiades, Coma Berenices, Hyades open cluster as a function of $T_{\rm{eff}}$. Data from the Ursa Major moving group are also depicted in this plot. The square symbols represent the early type A field stars. The open diamond symbols are members of NGC 3293. \textit{The size of the symbols is inversely proportional to $\log g$ ranging between 1.0 and 5.0 dex. The maximum in $\xi_t$ is observed around 8000 - 9000 K.}\\
In the right panel of Fig.~1 we add the results of Takeda et al. (2008) (in triangle symbols). Takeda et al. (2008) derived an analytical relation between $\xi_t$ and $T_{\rm{eff}}$ for 46 A-type field stars 
%given by 
%$$\xi_t^{std}=4 \rm{exp}[-\log (\dfrac{T_{\rm{eff}}}{8000})/A^2] \ \ \rm{with} \ \ A=(\log \dfrac{10000}{8000}) \dfrac{1}{\sqrt{ln 2}}$$
with an uncertainty of $\pm$30\% represented by the dotted curve line. In the same manner, Takeda et al. (2012) derived an analytical relation for $\xi_t$ for F and G stars represented in dashed line for a surface gravity of $\log g$= 4.5 dex.  \\
We also show the results of Edvardsson et al. (1993) pertaining to F-G stars with temperatures between 5550 and 6800 K. The analytical relation derived in their paper and plotted here in long-dashed line for a surface gravity of $\log g$= 4.5 dex. Bruntt et al. (2012) derived a new calibration for the microturbulent velocity versus the effective temperature for F and G stars based on a study of 93 solar-type stars that are targets of the NASA/Kepler mission. Their results are shown a $\log g$ of 4.5 dex (dashed-dotted line).

\begin{center}
\begin{figure}[htb!]
\includegraphics[scale=0.3]{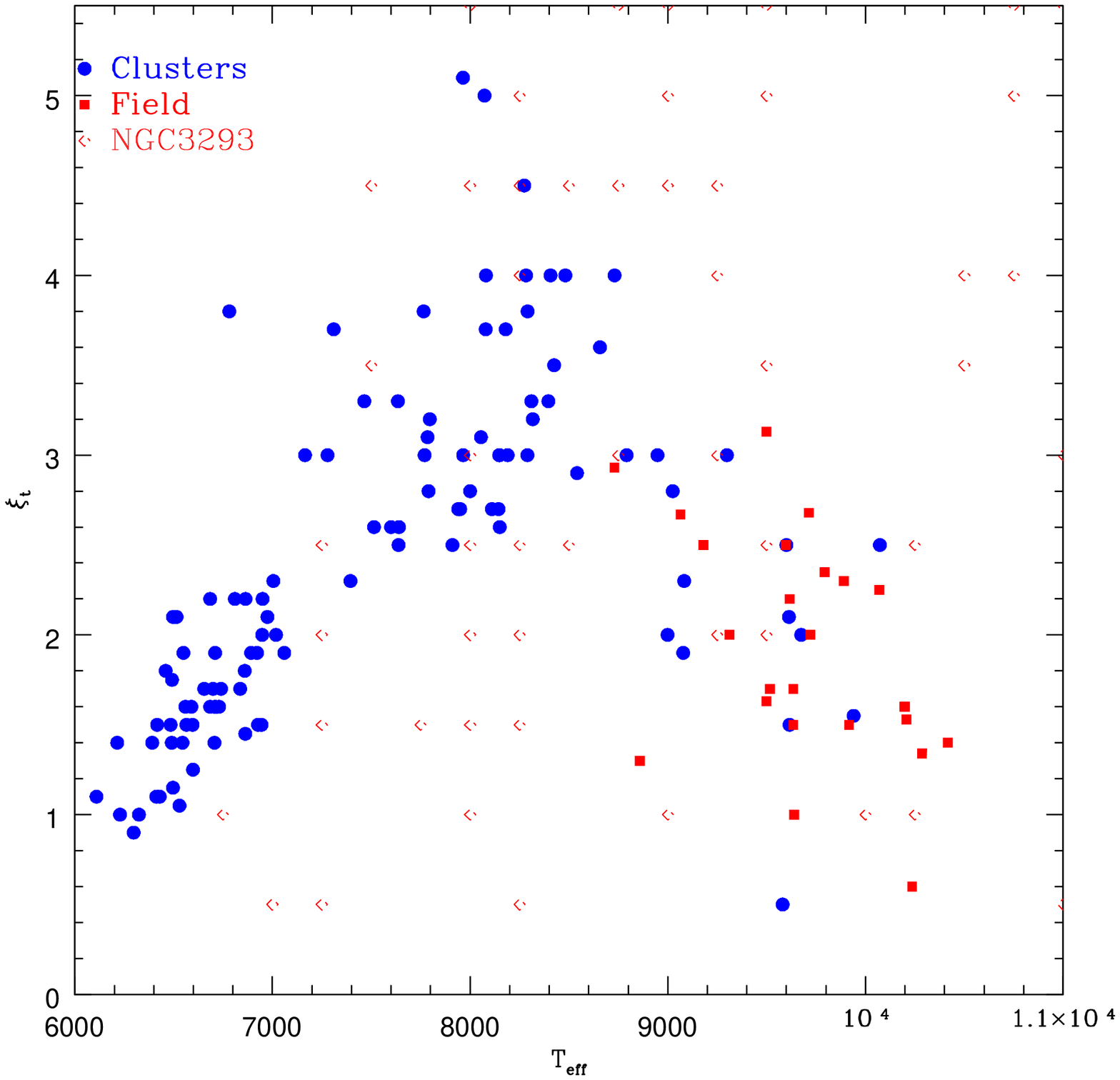}
\includegraphics[scale=0.3]{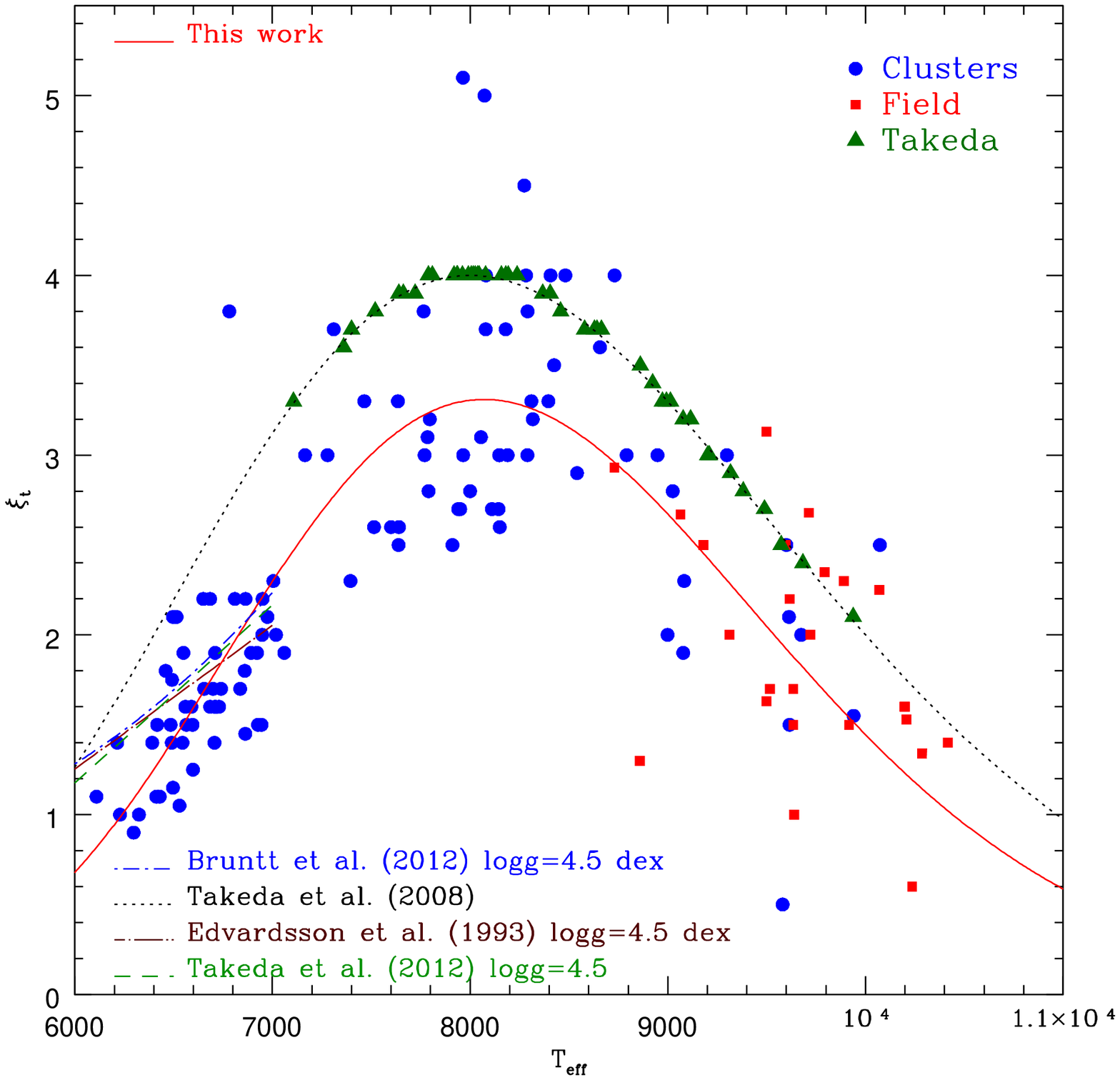}
\caption{Left: microturbulent velocity in km/s as a function of effective temperature in K for the A and F stars (circles) analysed in this work. Diamond symbols are members of the young open cluster NGC 3293 observed for the GES. The errorbars of the GES data are $\pm$250 K in $T_{\rm{eff}}$, and $\pm$0.5 in $\xi_t$. Right: we add the sample stars in triangle and the analytical formula (dotted line) of Takeda et al. (2008). The dashed-dotted line is the analytical relation of Bruntt et al. (2012) for a surface gravity of $\log g= 4.5$ dex. The dashed lines is the one of Takeda et al. (2012). The long-dashed line represents the relation given in Edvardsson et al. (1993) for a $\log g$ of 4.5 dex. }
\end{figure}
\end{center}
\section{Conclusions}

Inspection of Fig. 2 shows that $\xi_t$ reaches a broad maximum of 4$\pm$1 km/s around 8000 K and then drop tp 1 km/s at higher and cooler temperatures. This result has been confirmed by Takeda et al. (2008) and Smalley's (2004) prescriptions on convection in tepid stars.\\
We have looked for analytical function that would fit the observed shape over the range of A and F stars (fromm 6000 to about 10000 K). The relation that we found is given by

$$\xi_t = 3.31 \times \rm{exp}[-(\log (\frac{T_{\rm{eff}}}{8071.03})^2/0.01045)]$$

with an error of $\pm$25 \% shown by the continuous red curve.\\
Compared to Takeda et al. (2008) results, we can deduce a large dispersion of our findings around the analytical curve of their paper for A stars.
% by 
%$$\xi_t^{std}=4 \times \rm{exp}[-(\log (\frac{T_{\rm{eff}}}{8000})/A)^2] \ \ \ \rm{with} \ \ A=(\log \frac{10000}{8000}) \frac{1}{\sqrt{ln 2}}$$

One should note that the fit to our results (red curve) actually comes
close to Takeda's lower limit. Our microturbulent velocities are about
lower than his by about 1 km/s. \\
Concerning the F stars, the comparisons with the analytical relations of Edvardsson et al. (1993), Bruntt et al. (2012), and Takeda et al. (2012) 

%given by 
%$\xi_t = 1.25 + 8 \times 10^{-4} (T_{\rm{eff}}-6000) - 1.3 (\log g -4.5) \ \rm{km/s}$, of Takeda et al. (2012) given by $\xi_t = 9.96 \times 10^{-4} T_{\rm{eff}} -0.41 \times \log g -2.92 \ \rm{km/s}$, and the relation of Bruntt et al. (2012) given by $\xi_t = 1.095 + 5.44 \times 10^{-4} (T_{\rm{eff}}-5700) + 2.56 \times 10^{-7} (T_{\rm{eff}}-5700)^2 - 0.378 (\log g -4.0) \ \rm{km/s}$
show a good agreement for stars with $\log g$ larger than 4.0 dex. Our observed F stars have surface gravities between 4.1 and 4.6 dex. It follows that the best fit occurs with the curve of Edvardsson et al. (1993) that has a $\log g$ of 4.5 dex. The same applies to Takeda et al. (2012) and Bruntt et al. (2012) results.

\bigskip
{\it Acknowledgements.}  We made use of the NIST, SIMBAD, and VALD databases. A.L. acknowledges funding from the ESA/Belgian Federal
Science Policy in the framework of the PRODEX programme.